\documentclass[12pt,times, letterpaper] {article}
\usepackage{graphicx, indentfirst, setspace}
\usepackage[margin=1 in] {geometry}

\newcommand{\ipcomp}{I_{\rm pl-bin}}
\newcommand{\ikoz}{I_{\rm Koz}}
\newcommand{\tkoz}{\tau_{\rm Koz}}
\newcommand{\twkoze}{\tau_{\rm Koz,\oplus}}
\newcommand{\twkozj}{\tau_{\rm Koz,Jup}}
\newcommand{\tnkoze}{\tau_{\rm \Omega,Koz,\oplus}}
\newcommand{\tnkozj}{\tau_{\rm \Omega,Koz,Jup}}
\newcommand{\twppe}{\tau_{\rm  pp,\oplus}}
\newcommand{\twppj}{\tau_{\rm  pp,Jup}}
\newcommand{\tnpp}{\tau_{\rm \Omega,pp}}
\newcommand{\eearth}{e_\oplus}
\newcommand{\ajup}{a_{\rm Jup}}

\title{Pervasive Orbital Eccentricities Dictate the Habitability of Extrasolar Earths} 
\author
{Ryosuke Kita,$^{1}$ Frederic Rasio,$^{1,2}$ and Genya Takeda$^{1}$\\
\\
\normalsize{$^{1}$Department of Physics and Astronomy, Northwestern University,}\\
\normalsize{$^{2}$Center for Interdisciplinary Exploration and Research in Astrophysics,}\\
\normalsize{(CIERA), Northwestern University,}\\
\\
\normalsize{2145 Sheridan Road, Evanston, IL 60208, USA}\\
\\
\normalsize{Submitted to Astrobiology}
}

\date{}

\begin{document}

\maketitle{}

\section{Abstract}
The long-term habitability of Earth-like planets requires low orbital eccentricities.  A secular perturbation from a distant stellar companion is a very important mechanism in exciting planetary eccentricities, as many of the extrasolar planetary systems are associated with stellar companions.  Although the orbital evolution of an Earth-like planet in a stellar binary is well understood, the effect of a binary perturbation to a more realistic system containing additional gas giant planets has been very little studied.  Here we provide analytic criteria confirmed by a large ensemble of numerical integrations that identify the initial orbital parameters leading to eccentric orbits.  We show that an extra-solar earth is likely to experience a broad range of orbital evolution dictated by the location of a gas-giant planet, necessitating more focused studies on the effect of eccentricity on the potential for life.  

\section{Introduction}
\subsection{Habitability}
   Habitability of a planet is conventionally defined as the capability for liquid water to be sustained on a planetary surface.  A high (or low) level of stellar flux incident upon a planetary surface results in the loss of liquid water through evaporation by the runaway-greenhouse effect (or 
through freezing by global refrigeration) (Kasting \textit{et al.}, 1993).  Planets with eccentric orbits experience varying levels of stellar flux throughout a year as the distance from the star fluctuates during an orbit.  The time-averaged stellar flux $<F>$ over the orbital period is 
 \begin{equation}
 <F>={L\over 4\pi a^2 (1-e^2)^{1/2} },
 \end{equation} 
where $L$  is the luminosity of the host star, and $a$ and $e$ are the semi-major axis and eccentricity of the planet, respectively (Williams and Pollard, 2002).  As seen in Eq. 1, the eccentricity growth of a planet results in an increase in the time-averaged stellar flux and thus eventually leads to unhabitable conditions.  Climate models show that the Earth would start to lose its surface water with a hypothetical eccentricity of 0.4 around the Sun;  if the eccentricity exceeds 0.7, the Earth would lose all of its liquid water through the runaway-greenhouse effect (Williams and Pollard, 2002).  Thus, a perturbation that causes a planet to deviate from a circular orbit can greatly impact its long-term climatic stability.  This in turn could disturb the possible origin, evolution, and prevalence of life on the planet.

\subsection{Kozai Mechanism}
A secular perturbation from a stellar companion is one of the most efficient mechanisms in exciting planetary eccentricities to very large values.  Unlike our own Solar System, at least 20\,\% of the $\sim240$ extrasolar planetary systems detected as of 2007 are members of multiple-star systems (Raghavan \textit{et al.}, 2006; Desidera and Barbieri, 2007; Eggenberger \textit{et al.}, 2007).  Moreover, the multiplicity among the current sample of planet-hosting stars should be higher than 20\,\% as the photometric searches for stellar companions around known planetary systems are still ongoing.  
 
These stellar companions, despite their large distances from the planetary systems (typical binary separations range from $\sim 10^2 $ -- $10^4\,$AU), can still secularly perturb the planetary orbits around the primaries through a unique three-body interaction called the ``Kozai mechanism'', in which the cyclic angular momentum exchange between a binary companion and a planet results in a large-amplitude eccentricity oscillation of the planet  (Kozai, 1962; Holman \textit{et al.}, 1997; Innanen \textit{et al.}, 1997).  The Kozai mechanism takes place when the initial relative inclination $\ipcomp$ between the planetary and binary orbits exceeds the critical Kozai angle  $\ikoz = 39.23^{\circ}$.  Under such an initial configuration, the planet's eccentricity grows from $\sim0$ and oscillates with an amplitude that is constant to the lowest order as
\begin{equation}
e_{\rm max} \simeq \sqrt{1-{5 \over 3} \cos^2{\ipcomp}}.
\end{equation} 
(Holman \textit{et al.}, 1997).
For example, an Earth-mass planet at 1\,AU would reach the habitability limits of $e =$0.4 and 0.7 if there is a stellar companion initially inclined by $\ipcomp = 45^\circ$ and $56^\circ$, respectively.  Note that the distribution of $ \ipcomp$ in space is most likely isotropic because the orbital orientation of binaries with separations greater than $\sim 100\,$AU are not expected to be correlated with the invariable plane of the planetary systems around the primaries (Hale, 1994; Takeda \textit{et al.}, 2008). 

During Kozai cycles, a stellar companion repeatedly applies a small torque on the planetary orbit.  This torque accumulated over many binary orbits results in the precession of the planet's pericenter argument and hence the oscillation of the orbital eccentricity.  The timescale of the pericenter and eccentricity evolution can be analytically estimated as 
\begin{equation}
 \tkoz \approx {2\over3\pi}{P_{\rm bin}^{2}\over P_1}{m_0+m_1+m_{\rm bin} \over m_{\rm bin}}(1-e_{\rm bin}^2)^{3/2}
 \label{tkoz}
\end{equation} 
(Kiseleva \textit{et al.}, 1998; Ford \textit{et al.}, 2000), where the subscripts $bin$, $0$, and $1$ refer to the binary companion, the primary star, and the planet, respectively.  If there is another source of perturbation that precesses the planetary orbit on the timescale shorter than $\tkoz$ given in Eq.~\ref{tkoz}, then the torque applied from the stellar companion to the planetary orbit secularly averages out to zero, and no eccentricity growth by the companion would be observed (Wu and Murray, 2003).

The Kozai timescale may largely vary depending on the system parameters; for example, an Earth-size planet with a solar-mass stellar companion at 750\,AU would experience Kozai oscillations with a period $\tkoz \approx 320\,$Myr.  If a stellar companion is closer, then the oscillation timescale would be largely reduced;  for example at 250\,AU, $ \tkoz \approx 15\,$Myr.  Note that this oscillation period is also relevant to the evolution of life on a planet.  For example, if the Earth had an eccentricity of 0.4, even though some liquid water would remain on the surface, the surface temperatures would exceed $70^{\circ}\,C$ near the pericenter of the orbit (Williams and Pollard, 2002).  At this temperature, thermophiles and other simple organisms may be able to comfortably survive, but complex life could be eliminated (Levy and Miller, 1998; Daniel and Cowan, 2000).  However, if the eccentricity oscillation period is sufficiently long (on the order of hundred million years), the low eccentricity duration of the cycle may provide an intriguing avenue for the evolution of more complex organisms.

Although the eccentricity variation of a planet in a three-body system is understood well, in reality, a single Earth-like planet in a binary is probably not a common configuration.  Both observations and numerical planet-formation simulations rather suggest that planetary systems naturally form in multiple configurations, containing one or a few gas giant planets  (Ida and Lin, 2005; Thommes \textit{et al.}, 2008).  Thus, it is certainly relevant to discuss the dynamical evolution of a habitable Earth-like planet, secularly perturbed by other planets and a distant stellar companion.  However, the evolution of gravitationally-coupled multiple planets under the influence of a secular binary perturbation is significantly more complex than the simple three-body Kozai mechanism and has been very little understood in previous studies.

\section{Methodology}
\subsection{Numerical Simulations}
Here we investigate a prototypical four-body system, containing the Sun, an Earth-mass planet, a Jupiter-mass planet, and a hypothetical stellar companion.  To determine the dependence of the earth's eccentricity on the orbital elements of the jupiter and the stellar companion, we have performed  an extensive set of numerical simulations of four-body systems with different initial binary parameters, $\ipcomp$ and $a_{\rm bin}$, as well as the initial location of the jupiter $a_{\rm Jup}$.  To ensure that there is no immediate instability in the system, we located  the jupiter safely within the Hill stable limits for all the ensembles (Gladman, 1993).  

The orbital evolution of the four-body system was followed by integrating the full equations of motion with the Bulirsch-Stoer integration scheme (Chambers, 1999).  We have also included the post-Newtonian correction to observe the general relativistic (GR) effect on the earth that occasionally competes with the secular perturbations from the other bodies.  After numerically integrating each system for 500\,Myr, we have recorded the maximum eccentricity attained by the earth during the evolution.  

The post-Newtonian corrections are relevant in these simulations because a wide binary companion's perturbation can be weaker than the GR precession of the earth.  For a Sun/Earth system with a solar-mass binary companion, the GR precession overtakes the binary perturbation when the semimajor axis of the companion is greater than $\sim 500 AU$.  Note that in our simulations this situation does not ensure a stable circular orbit for the earth, because of the presence of another planet.

The 500\,Myr simulation is sufficiently long enough to catch all of the systems that would result in an eccentric earth.  500\,Myr is longer than the characteristic perturbation timescales derived from the Kozai mechanism or the Laplace-Lagrange secular theory.  For the eccentric cases, the maximum eccentricity of the earth was reached quickly within 50 Myr, while the stable-circular orbits showed no indication of ever becoming eccentric. 

Each system was integrated 10 times with random initial phases to confirm that the initial phase does not affect the overall behavior. The maximum eccentricity of the earth that resulted was very similar for each initial phase due to the secular nature of the dynamical behaviors.  For Figure 1, for each system, the average of the maximum eccentricities of the 10 simulations was used.

Figure 1 illustrates how the earth's maximum eccentricity depends on $\ajup$ and $\ipcomp$, with the binary separation fixed at 750\,AU.  We first note that without the presence of the jupiter, this map would look completely dark; the stellar companion perturbation on the earth is too weak - the Kozai oscillations ($\tau_{Koz} \approx 100\, \rm Myr$)  are slow compared to the GR precession ($\tau_{GR} \approx 30\, \rm Myr$) -  
such that the orbit of the earth would remain nearly circular, no matter how large the initial inclination is.  If the companion perturbation was stronger than GR ($\tau_{Koz} < \tau_{GR}$), without the jupiter, the earth would have experienced eccentricity oscillations with amplitudes determined by $\ipcomp$, above $\ikoz$.  With the presence of a jupiter, however, the earth's orbital evolution is not simply constrained by $\ipcomp$ or the strength of the companion perturbation on the earth.  Rather, the location of the jupiter and its consequent orbital evolution sensitively affects the earth's eccentricity and its potential for life.

\subsection{Timescale Analysis}
This complex parameter dependence exhibited in Figure 1 can be understood by comparing, on both planets, the strength of the binary perturbation with that of the mutual perturbation the planets apply onto each other (hereafter denoted with subscripts ``koz'' and ``pp'', respectively).  Each of these perturbations can be quantified by their precession timescales $\tau$ of the argument of pericenter $\omega$ and ascending node $\Omega$.  The binary perturbation and planet perturbation timescales were derived from the Kozai perturbation theory and the Laplace--Lagrange secular theory, respectively  (Takeda \textit{et al.}, 2008).  To distinguish between the perturbations on the earth and jupiter, we denote the timescales with subscripts $\oplus$ and ``Jup'', respectively.  Figure 2 shows the timescales of the various perturbations computed over the same simulation space shown in Figure 1.  This timescale analysis is applicable when the binary orbit is relatively wide compared to the planetary system (to apply the Kozai timescale estimate), and when the planets are roughly coplanar (to apply the Laplace-Lagrange timescale estimate) (Takeda \textit{et al.}, 2008).

\section{Results} 
\subsection{Analysis of Simulation Results}
Here we will explore the parameter space of Figure 1 and 2, using the timescale analysis to justify the observations in the numerical simulations.  

Figures 3 - 7 are typical simulation results from each of the dynamical regions to be described.  Throughout these figures, the lighter marks indicate the jupiter's parameters, while the heavy marks indicates the earth's parameters.  The figures show how the planets' ascending node $\Omega$, argument of pericenter $\omega$, inclination $i$, and eccentricity $e$ change over time, from top to bottom, respectively.  The inclination is measured using the binary plane as the reference plane.  The dotted line in the inclination frame shows the mutual inclination between the planets; because the ascending nodes of the planets precess, the difference between the inclinations of the planets relative to the binary plane does not indicate the planet's mutual relative inclination.  

When the jupiter is sufficiently close to the earth but safely outside the Hill stability boundary ($1.5\,$AU $< \ajup < 3\,$AU; Region\,A in Figure 1), for both planets, their respective planet-planet mutual perturbations dominate over the binary perturbation ($\twppj < \twkozj$ , $\twppe < \twkoze$).  As a result, even if the planets are initially inclined with respect to the binary orbit by more than $\ikoz$, the Kozai mechanism is completely suppressed, and neither planet undergoes eccentricity oscillations in this region.

When the jupiter's initial location is moved further from the earth, the jupiter undergoes the Kozai mechanism because the perturbation from the 
binary companion begins to dominate over the perturbation caused by the earth ($\twkozj < \twppj$, Region\,B and C).  In Region\,B, $\ipcomp$ is initially below $\ikoz$ and thus the orbits of jupiter and the earth remain nearly circular.  In Region\,C ($3 \,$ AU $ < \ajup < 13 \,$ AU, 
$\ipcomp > 40^\circ$), an interesting coupling behavior between the earth and jupiter is observed.  The jupiter is subject to the Kozai mechanism caused by the stellar companion; its orbital inclination and eccentricity oscillate on a timescale $\twkozj$.  In the simulations, we also observed the earth's eccentricity and inclination oscillate in concert with those of the jupiter, also on the timescale $\twkozj$.  This is a result of the gravitational coupling between the planets; specifically, the strength of the nodal precession by the planet-planet perturbation ($ \tnpp< \tnkozj, \tnkoze $) forces the two planetary orbits to precess together, while maintaining mutually coplanar orbits.  Once $\ipcomp$ exceeds $70^\circ$, the large orbital eccentricity on the jupiter induced by the binary companion results in the two planetary orbits crossing.  Crossing orbits typically scatter the earth off its original orbit to an orbit with an eccentricity much beyond the habitable limit.

When the orbit of the jupiter is further widened to the range 13\,--\,25\,AU, another dynamically interesting Region\,D arises.  Notice that in this region the earth gains significantly large orbital eccentricity, even though initially the binary and the planetary orbits are below the critical Kozai angle ($\ipcomp < \ikoz$).  Because the earth and jupiter are widely separated, the planet-planet perturbation is not sufficiently strong to maintain the coplanar precession observed in Region B and C ($\tnkozj < \tnpp$).  As a result, the earth's nodal precession lags behind that of the jupiter, generating a nodal offset angle $\Delta\Omega$ between the two orbits.  To compensate, the nodal precession of the earth begins to accelerate, effectively minimizing $\Delta\Omega$.  Nodal acceleration of the earth results in a increase in its orbital inclination to conserve the component of the orbital angular momentum normal to the binary orbit.  The growth of the earth's orbital inclination with respect to the binary orbit also generates a large mutual inclination angle between the two planetary orbits, beyond $\ikoz$.  As a result, the jupiter begins to apply a Kozai-type perturbation on the earth, thus raising $\eearth$ to large values without the need of an initially high $\ipcomp$.  Although GR precession can suppress such regions of high eccentricity, if the outer planet was a gas giant, this will only occur in extreme cases.    

The nodal precession kicks by the jupiter and the subsequent additional Kozai cycles of the earth gradually disappear as the jupiter's perturbation further weakens ($\ajup > 25\,AU$).  As the planet-planet interaction weakens, the earth's orbit becomes stable and circular because of GR precession or experiences Kozai cycles induced by the binary perturbation (when $\ipcomp > \ikoz$).  Region E in Figure 1 shows an earth with a circular orbit for most of the systems.  High eccentricities would be seen in this region above $\ikoz$, if the GR precession did not suppress the binary perturbation.    

\subsection{Comparison of Timescale analysis}
Following the timescale analysis, we expect the maximum orbital eccentricities of an earth to maintain a similar distribution as the parameters of the outer planet and companion are varied.  Figure 8 compares the numerical simulations with the timescales for three sets of simulations conducted with different binary separations. The numerical boundaries between different dynamical classes match  well with the analytically estimated boundaries yielded from the timescale comparisons.

 \subsection{Role of General Relativity}
GR precession competes with all of the perturbations that affect the earth.  In Figure 1, GR has little effect because it is not the dominant interaction throughout most of the simulation space.  This is due to the small mass of the earth and sun, in addition to the relatively large distance between the bodies.  Yet, as parameters change, such as the mass of the sun, the semimajor axis of the earth, and the mass of the earth, stronger GR precession may play a noticeable role in the orbital dynamics of the earth.  However, it is difficult to consider such situations in the context of habitability because these parameters play a direct role in the stellar flux and habitability of the earth, independent of eccentricity.  Thus, we only considered the possibility of GR suppression in the cases where the planet-planet or Kozai interactions become weaker, such as by a smaller, more distant binary companion or a gas-giant planet.  Figure 9 shows the outer planet parameters at which GR effects can suppress the eccentricities observed in the highly eccentric region D on Figure 1.  Although the GR precession can theoretically suppress such regions of high eccentricity, if the outer planet was a gas giant, this can only occur in extreme cases.     

\section{Discussion}

\subsection{Using the Timescale Analysis}	

The timescale analysis provides a method to quantify the strength of the dynamical interactions, thus creating the ability to predict the behaviors that should arise from a wide range of planetary systems hosted within a wide-binary.  The consistency between the behaviors seen in the numerical simulations with the behaviors predicted by the timescale analysis demonstrates the effectiveness of this approach, as shown in Figure 8.  

Using the timescale analysis, we can show, without exploring the parameter space with numerical simulations, that a different range of parameters would result in a similar set of orbital behaviors.  Figure 10 shows the parameters at which an earth-like planet will experience the high eccentricities observed in Region D.  The parameters were determined by obtaining the specific conditions determined by the timescale analysis.

\subsection{Distribution of Eccentricities}
Most notably, our simulations indicate that an earth is likely to exhibit high eccentricities across a wide range of parameters.  Figure 11 illustrates the probability of the earth reaching the habitable eccentricity limits of 0.4 and 0.7 as a function of the jupiter's semimajor axis.  The similarity between the dotted and solid lines at the dynamically rigid and nodal libration regions ($a_{\rm 2} < 20$) indicate that the Earth is likely to experience extremely high eccentricity oscillations rather than a distribution of high and moderate eccentricities like the single-planet case.  As shown with the timescale analysis and additional simulation sets, a similar distribution should be seen across a range of parameters.  Thus, an earth-like planet within a binary system will most likely exhibit convincingly non-habitable eccentricities.  However, it is important to note the length of time at which the planet experiences the high orbital eccentricities.  As described above, the earth-like planet may experience long periods of low eccentricities between its periods of high eccentricities.  Such secular behavior must be carefully studied in the context of the evolution of life before ruling out the possibility of a habitable earth.  

\subsection{Conclusion}
 Our results demonstrate the variety of orbits that an earth can exhibit within a binary system when accompanied by a second planet.  With such a high prevalence of binaries, in addition to the strong likelihood for multiple planets within a system, we should expect eccentric earths even when the companion is distant.  As the ability to search for extrasolar earths increases with observational advances, the possibility of other bodies within the system must also be taken into account when discussing habitability.  The diverse orbital behavior that result from the presence of such bodies highlights the necessity to study in detail extrasolar earths without stable circular orbits.  Thus, we call on climatologists, biologists, and astrobiologists to carefully consider the broad range of possible eccentricities when studying the potential for the origin, evolution, and prevalence of life on a planet.     

\section{Acknowledgements}
This work was supported by NSF Grant AST-0507727.  R.K. was supported by the NASA Summer Research Program for College Students and an Undergraduate Research Grant at Northwestern University. We thank Lamya Saleh for providing us an improved version of the Mercury BS integrator including the lowest-order post-Newtonian correction. 

\section{Author Disclosure Statement}
No competing financial interests exist.

\section{References}

Chambers, J.E. (1999) A hybrid symplectic integrator that permits close encounters between massive bodies. \textit{Monthly Notices of the Royal Astronomical Society} 304, 793-799. \\ \\
Daniel, R.M. and Cowan, D.A. (2000) Biomolecular stability and life at high temperatures. \textit{Cellular and Molecular Life Sciences}, 57, 250-264. \\ \\
Eggenberger, A., Udry, S., Chauvin, G., and Beuzit, J.L. (2007) The impact of stellar duplicity on planet occurrence and properties. I. Observational results of a VLT/NACO search for stellar companions to 130 nearby stars with and without planets. \textit{Astronomy and Astrophysics}, 474, 273-291. \\ \\
Desidera, S. and Barbieri, M. (2007) Properties of planets in binary systems. The role of binary separation. \textit{Astronomy and Astrophysics}, 462, 345-353. \\ \\
Ford, E.B., Kozinsky, B., and Rasio, F.A. (2000) Secular Evolution of Hierarchical Triple Star Systems. \textit{Astrophysical Journal}, 535, 385-401. \\ \\
Gladman, B. (1993) Dynamics of systems of two close planets. \textit{Icarus}, 106, 247-263.\\ \\
Hale, A. (1994) Orbital coplanarity in solar-type binary systems: Implications for planetary system formation and detection. \textit{Astronomical Journal}, 107, 306-332. \\ \\
Holman, M., Touma, J., and Tremaine, S. (1997) Chaotic variations in the eccentricity of the planet orbiting 16 Cygni B. \textit{Nature}, 386, 254-256. \\ \\
Ida, S. and Lin, D.N.C. (2005) Toward a Deterministic Model of Planetary Formation. III. Mass Distribution of Short-Period Planets around Stars of Various Masses. \textit{Astrophysical Journal}, 626, 1045-1060. \\ \\
Innanen, K.A.,  Zheng, J.Q., Mikkola, S., and Valtonen, M.J. (1997) The Kozai Mechanism and the Stability of Planetary Orbits in Binary Star Systems. \textit{Astronomical Journal}, 113, 1915-1919. \\ \\
Kasting, J.F., Whitmire, D.P., and Reynolds, R.T. (1993) Habitable Zones around Main Sequence Stars. \textit{Icarus}, 101, 108-128. \\ \\
Kiseleva, L.G., Eggleton, P.P., and Mikkola, S. (1998) Tidal friction in triple stars. \textit{Monthly Notices of the Royal Astronomical 
Society}, 300, 292-302. \\ \\
Kozai, Y. (1962) Secular perturbations of asteroids with high inclination and eccentricity. \textit{Astronomical Journal}, 67, 591-598. \\ \\
Levy, M. and Miller, S.L. (1998) The stability of the RNA bases: Implications for the origin of life. \textit{Proceedings of the National Academy of Sciences} 95, 7933-7938. \\ \\
Raghavan, D., Henry, T.J., Mason, B.D., and Subasavage, J.P. (2006) Two Suns in The Sky: Stellar Multiplicity in Exoplanet Systems.\textit{Astrophysical Journal}, 646, 523-542. \\  \\
Takeda, G., Kita, R., and Rasio, F.A. (2008) Planetary Systems in Binaries. I. Dynamical Classification. \textit{Astrophysical Journal}, 683, 1063-1075. \\ \\
Thommes, E.W., Bryden, G., Wu, Y., and Rasio, F.A. (2008) From Mean Motion Resonances to Scattered Planets: Producing the Solar System, Eccentric Exoplanets, and Late Heavy Bombardments. \textit{Astrophysical Journal}, 675, 1538-1548. \\ \\
Williams, D.M. and Pollard, D. (2002) Earth-like worlds on eccentric orbits: excursions beyond the habitable zone. \textit{International Journal of Astrobiology}, 1, 61-69. \\ \\
Wu, Y. and Murray, N. (2003) Planet Migration and Binary Companions: The Case of HD 80606b. \textit{Astrophysical Journal}, 589, 605-614. \\

\section{Figures}

\begin{figure}
\begin{center}
\includegraphics[width=180mm]{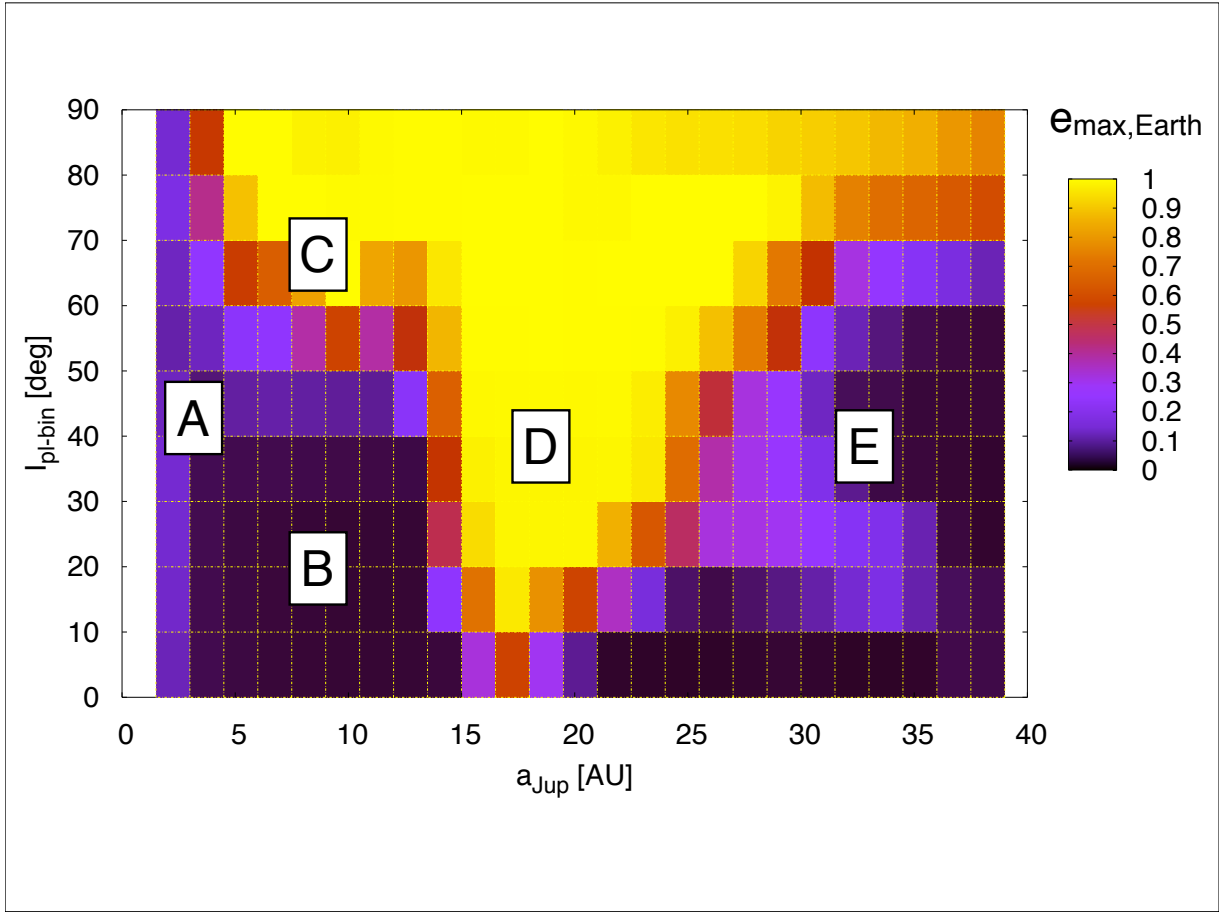}
\end{center}
\caption{Maximum orbital eccentricity of the earth during 500\,Myr of evolution for different sets of initial $\ipcomp$ and $\ajup$ with fixed initial binary parameters $a_{\rm bin} = 750\,$AU and $e_{\rm bin} =$ 0.2.    For the cases in which the earth collided with the central star, $e_{\rm max} = 1.0$ is recorded.  The map can be separated into five dynamically-unique regions marked by Region A to E.}
\end{figure}
\clearpage

\begin{figure}
\begin{center}
\includegraphics[width=150mm]{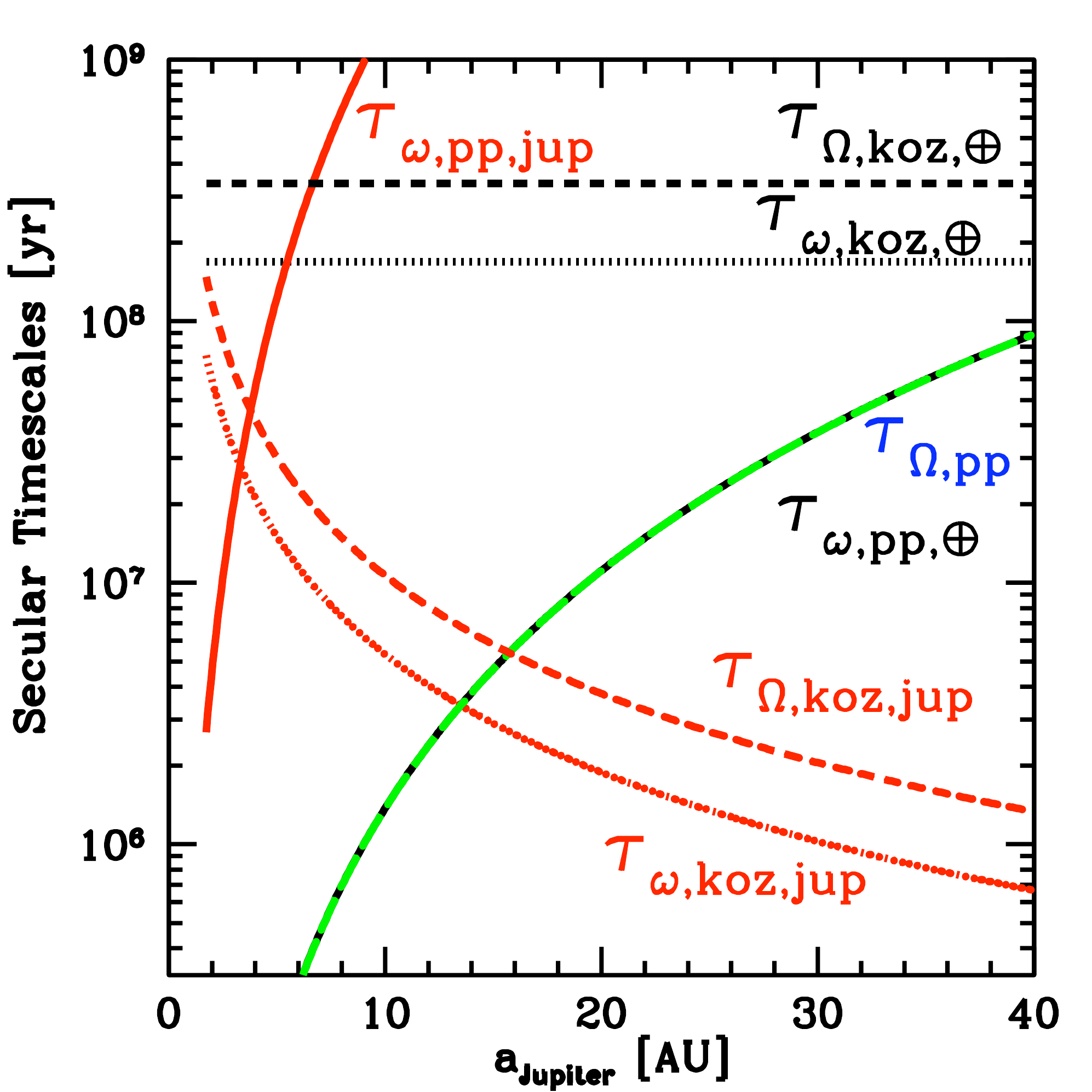}
\end{center}
\caption{ Secular orbital evolution timescales of the earth and jupiter in a binary as functions of $\ajup$.  The binary parameters are as in Figure 1.  A smaller timescale indicates a faster precession, which ultimately implies a stronger perturbation.  The Kozai oscillations can be suppressed if the planet-planet precession is faster than the Kozai precession.  A strong planet-planet nodal perturbation allows the planets' orbits to remain coplanar relative to each other throughout their evolution.}
\end{figure}
\clearpage

\begin{figure}
\begin{center}
\includegraphics[width=150mm]{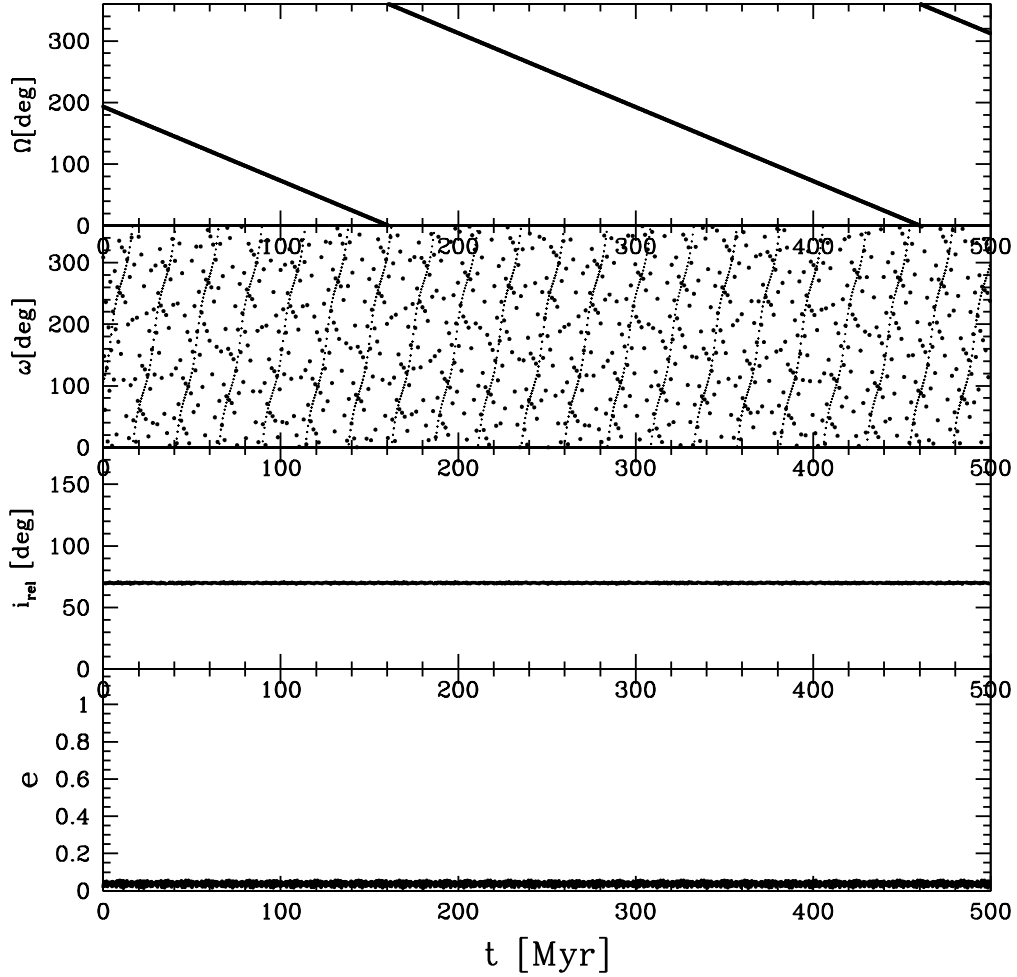}
\end{center}
\caption{  Simulation results of a representative system within Region A, where $\ipcomp$ is 70 degrees, and $\ajup$ is 3 AU.  Although the planets are both inclined over the Kozai angle, neither experience eccentricity oscillations because of the strong mutual planet-planet interaction.  The planets remain coplanar relative to each other.  The jupiter is circularized by the earth (notice the pericenter precession is $\sim$ $\twppj$), and the earth is circularized by the jupiter.   
}
\end{figure}
\clearpage

\begin{figure}
\begin{center}
\includegraphics[width=150mm]{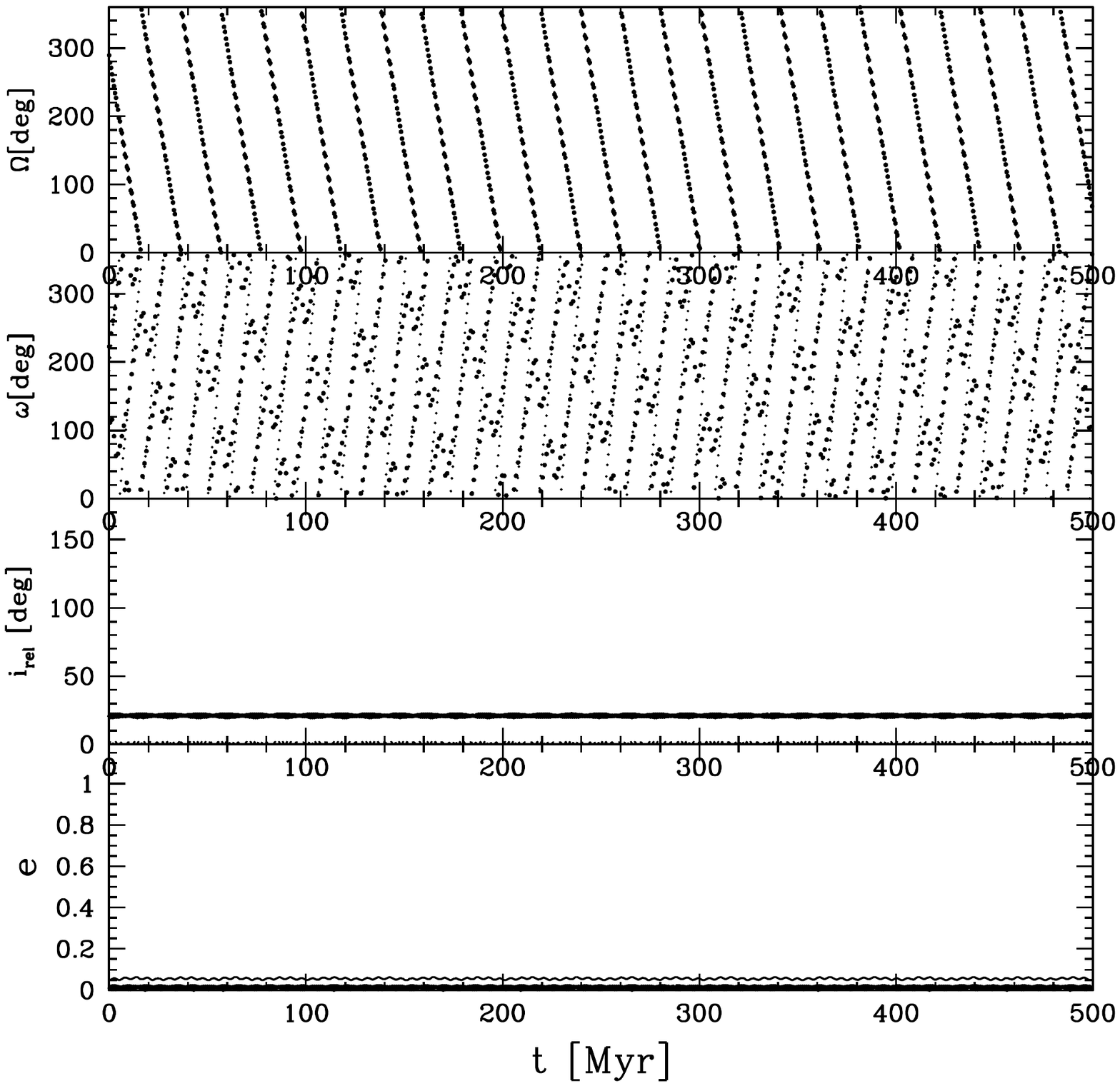}
\end{center}
\caption{ Simulation results of a representative system within Region B, where $\ipcomp$ is 20 degrees, and $\ajup$ is 9 AU.  The planets still remain coplanar, because of the strong $\tnpp$.  The jupiter, inclined below $\ikoz$, does not show the eccentricity oscillations although it experiences the binary perturbation as seen by its pericenter precession near $\twkozj$.  The earth remains circularized by the jupiter. }
\end{figure}
\clearpage

\begin{figure}
\begin{center}
\includegraphics[width=150mm]{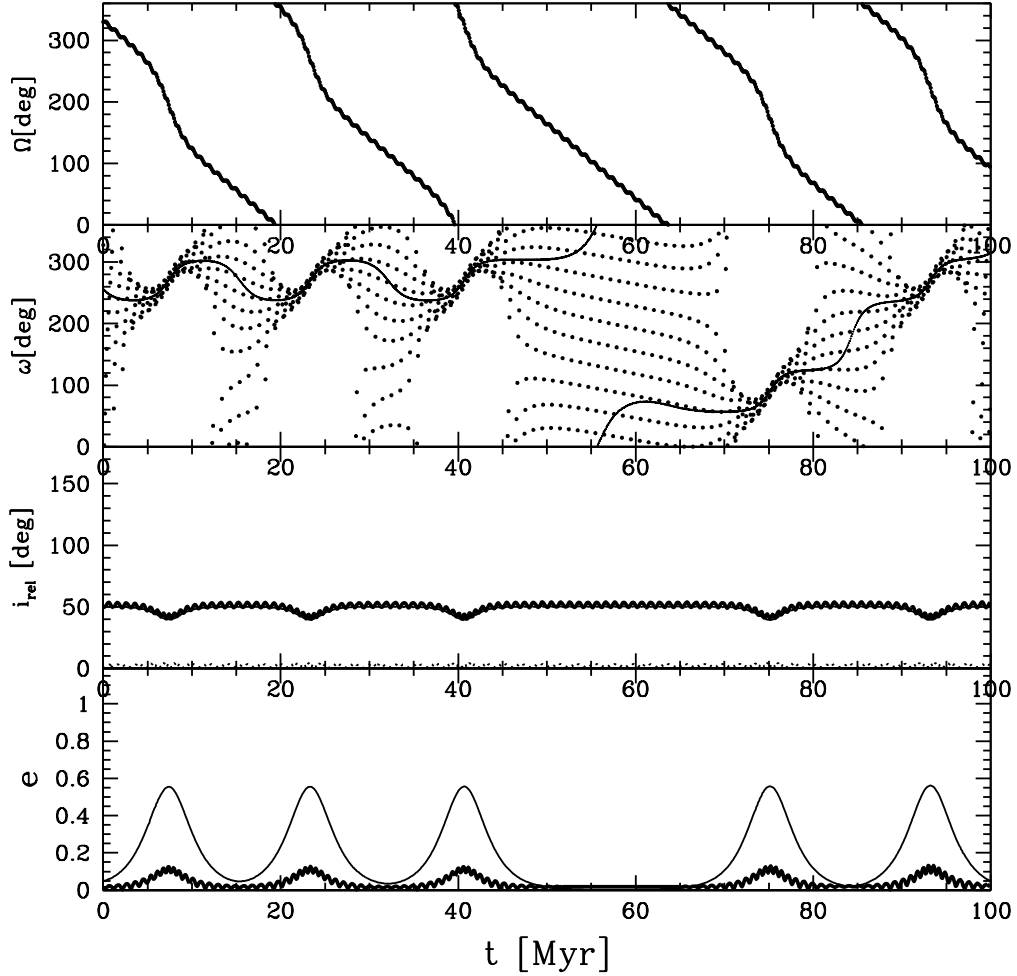}
\end{center}
\caption{ Simulation results depicting the concerted eccentricity described in Region C - in this particular system $\ipcomp$ is 50 degrees, and $\ajup$ is 9 AU.  Notice how throughout the eccentricity oscillations, the planets remain coplanar.   The jupiter, inclined over $\ikoz$, experiences the Kozai oscillations.  Because of the coplanarity maintained by the strong planet-planet interaction, the earth follows the jupiter's eccentricity oscillations.  
}
\end{figure}
\clearpage

\begin{figure}
\begin{center}
\includegraphics[width=80mm]{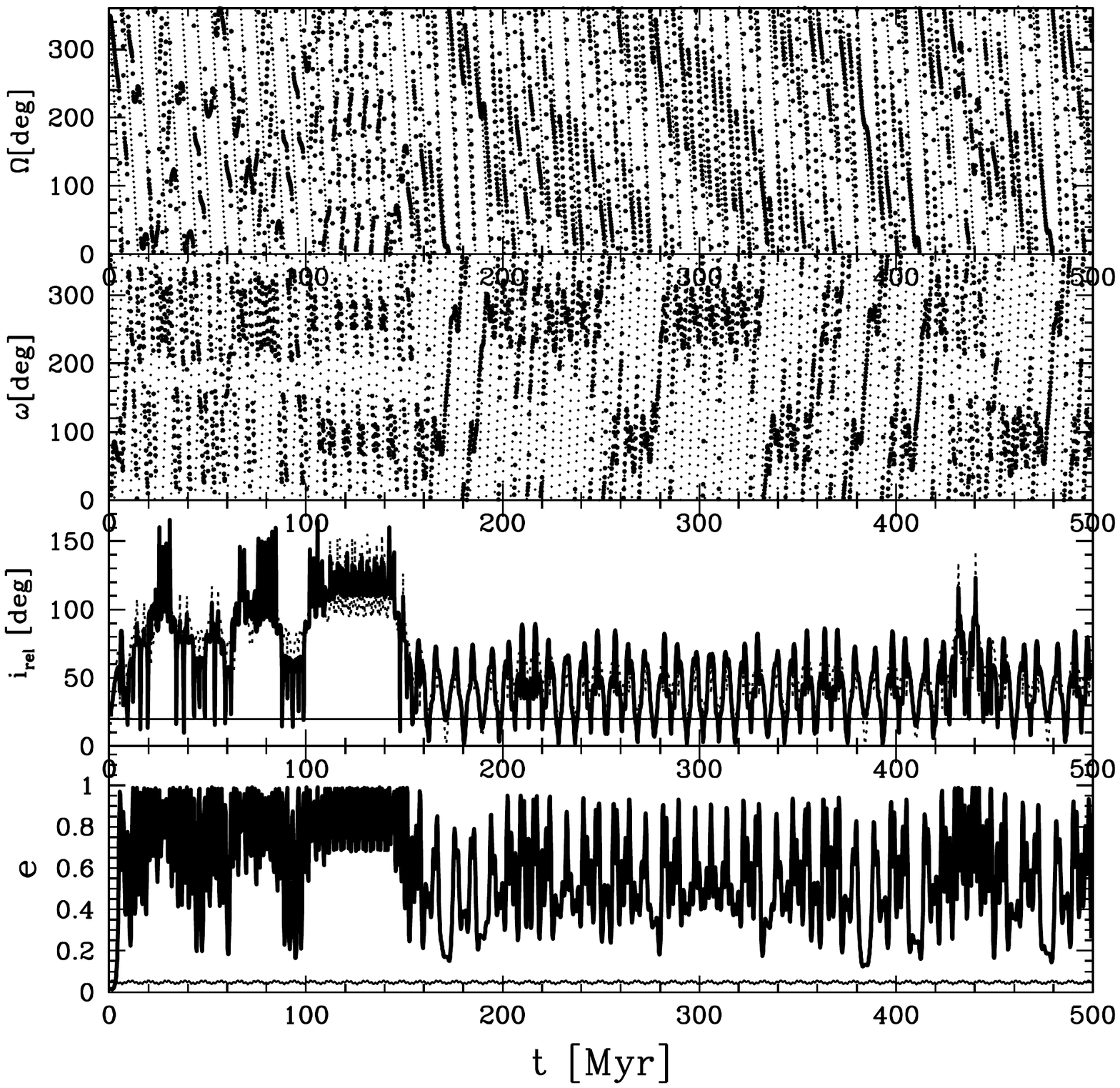}
\includegraphics[width=80mm]{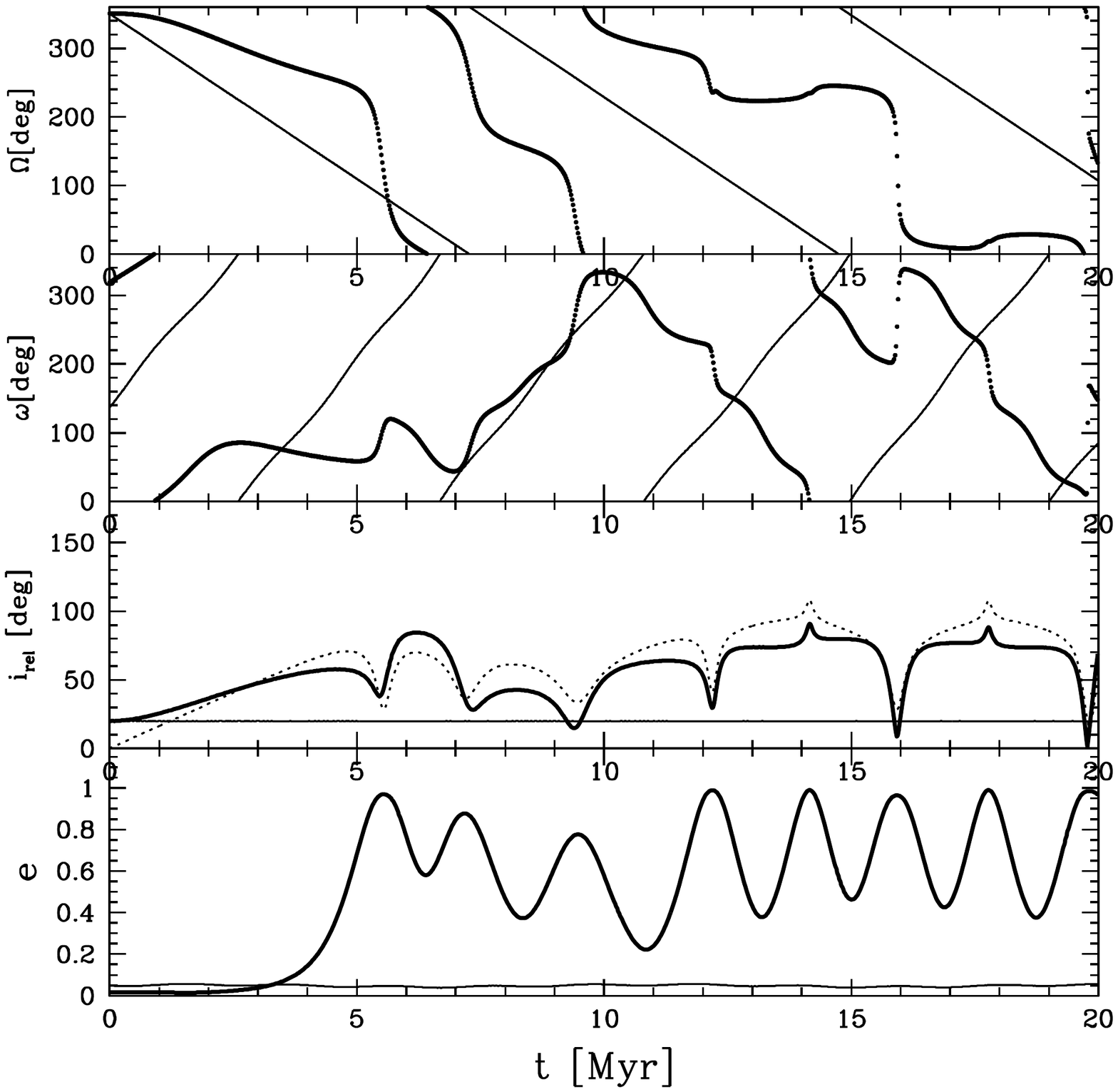}
\end{center}
\caption{ Simulation results of a representative system within Region D, where $\ipcomp$ is 20 degrees, and $\ajup$ is 18 AU. The right plot is the same simulation as the left, but shows only the first 20 Myr of the integration. The earth quickly experiences high eccentricities, although the planets are both initially inclined below the Kozai angle.  As the misalignment occurs in the ascending node, the mutual inclination between the planets  rises quickly over the Kozai angle.  This results in high eccentricities induced in the earth, which reappears often throughout the 500 Myr.}
\end{figure}
\clearpage

\begin{figure}
\begin{center}
\includegraphics[width=150mm]{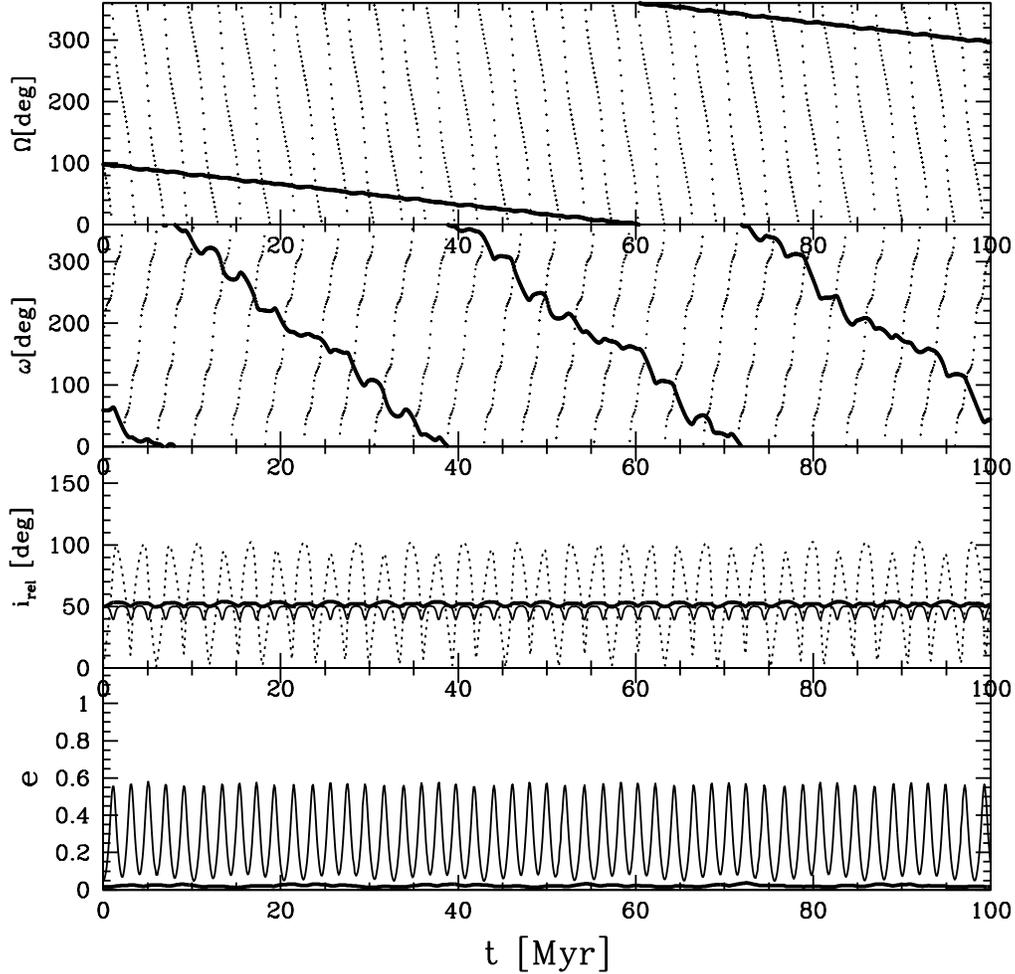}
\end{center}
\caption{  Simulation results of a representative system within Region E, where $\ipcomp$ is 50 degrees, and $\ajup$ is 31.5 AU.  The earth remains at low eccentricity throughout the 500 Myr because of the GR precession (pericenter precesses $\sim$ 30 Myr, the timescale of the GR precession).  The jupiter here experiences the Kozai oscillation, because the planets are initially inclined over $\ikoz$.  Although the planets become misaligned, and the mutual inclination becomes greater than the Kozai angle, the jupiter is distant enough from the earth, such that the GR precession suppresses the planet-planet interaction.  }
\end{figure}
\clearpage

\begin{figure}
\begin{center}
\includegraphics[width=150mm]{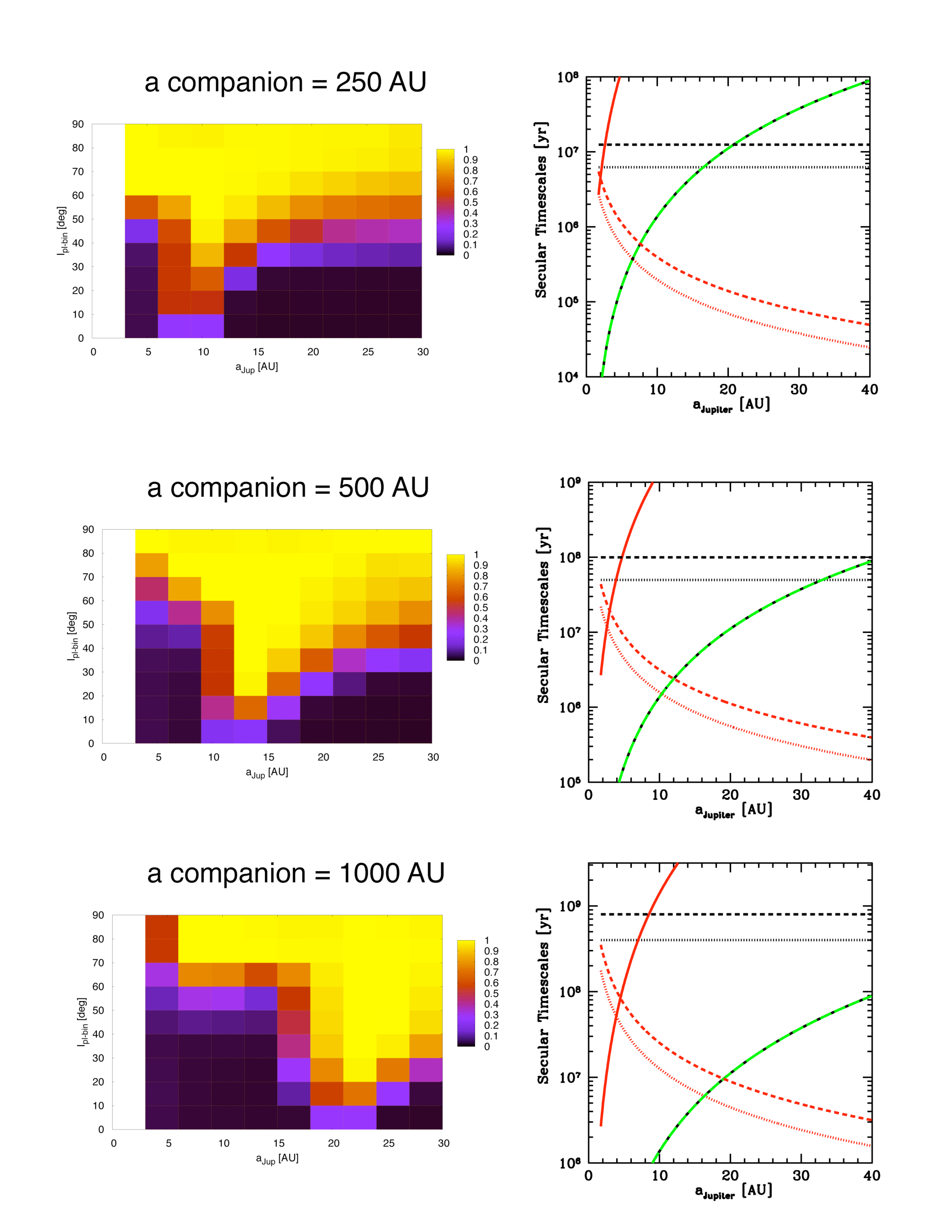}
\end{center}
\caption{ The consistency between the timescale analysis and the numerical simulations can be seen as the parameters for simulations sets are changed.  The three maps on the left are numerical simulations of a multiple-planet system with a binary companion, similar to the map described in Figure 1.  Each map has a different semimajor axis for the binary companion.  The timescale analyses on the right are derived from the Kozai mechanism and the Laplace-Lagrange secular theory - the type of line matches with the type shown in Figure 2.  
}
\end{figure}
\clearpage

\begin{figure}
\begin{center}
\includegraphics[width=150mm]{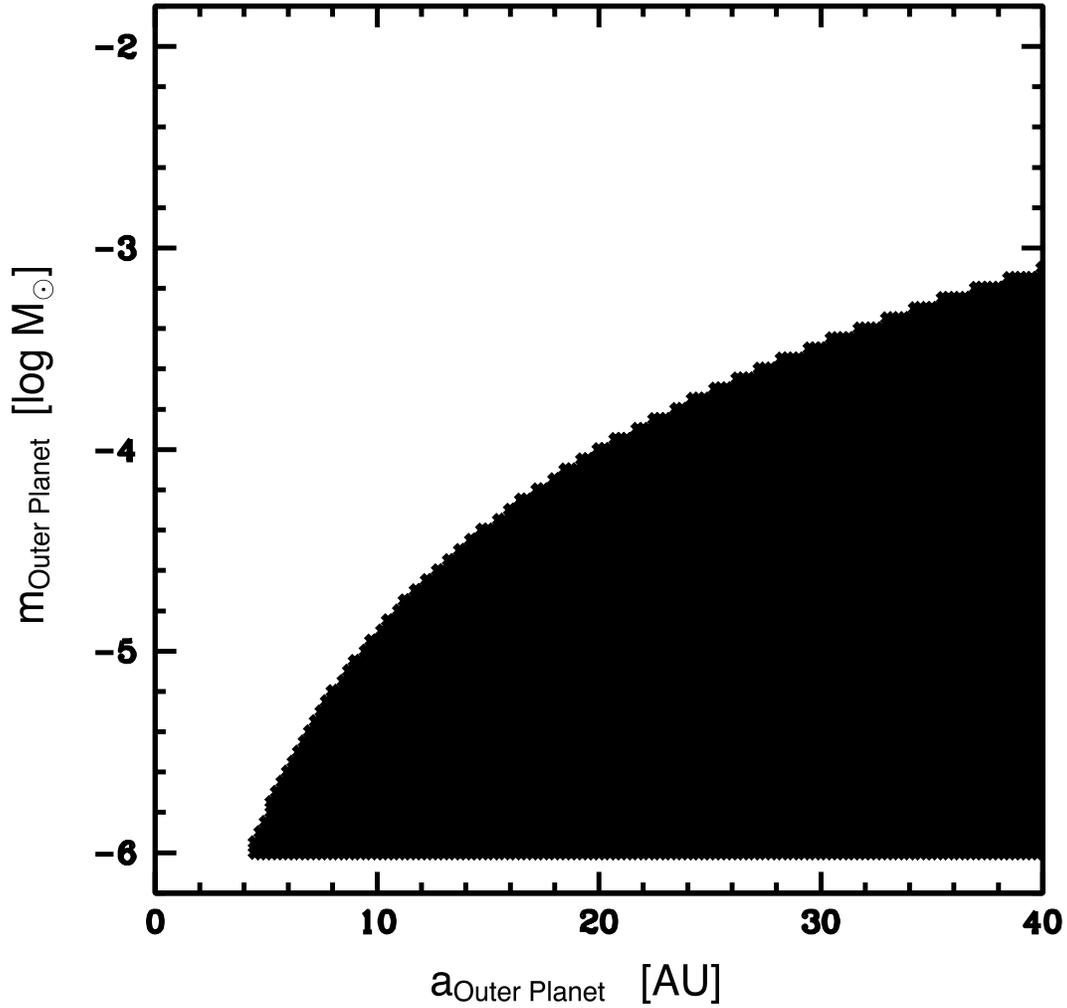}
\end{center}
\caption{ The outer planet parameters at which GR will suppress the Kozai cycles of an earth-like planet as described in Region D.  The ability for GR to suppress the eccentricities observed in Region D will depend on the eccentricity of the outer planet as well.  The shaded region depicts the largest possible range for the possibility of suppression.  Note how the high eccentricities observed in Region D will not be suppressed if the outer planet is at least as massive as Jupiter. 
}
\end{figure}    
\clearpage

\begin{figure}
\begin{center}
\includegraphics[width=150mm]{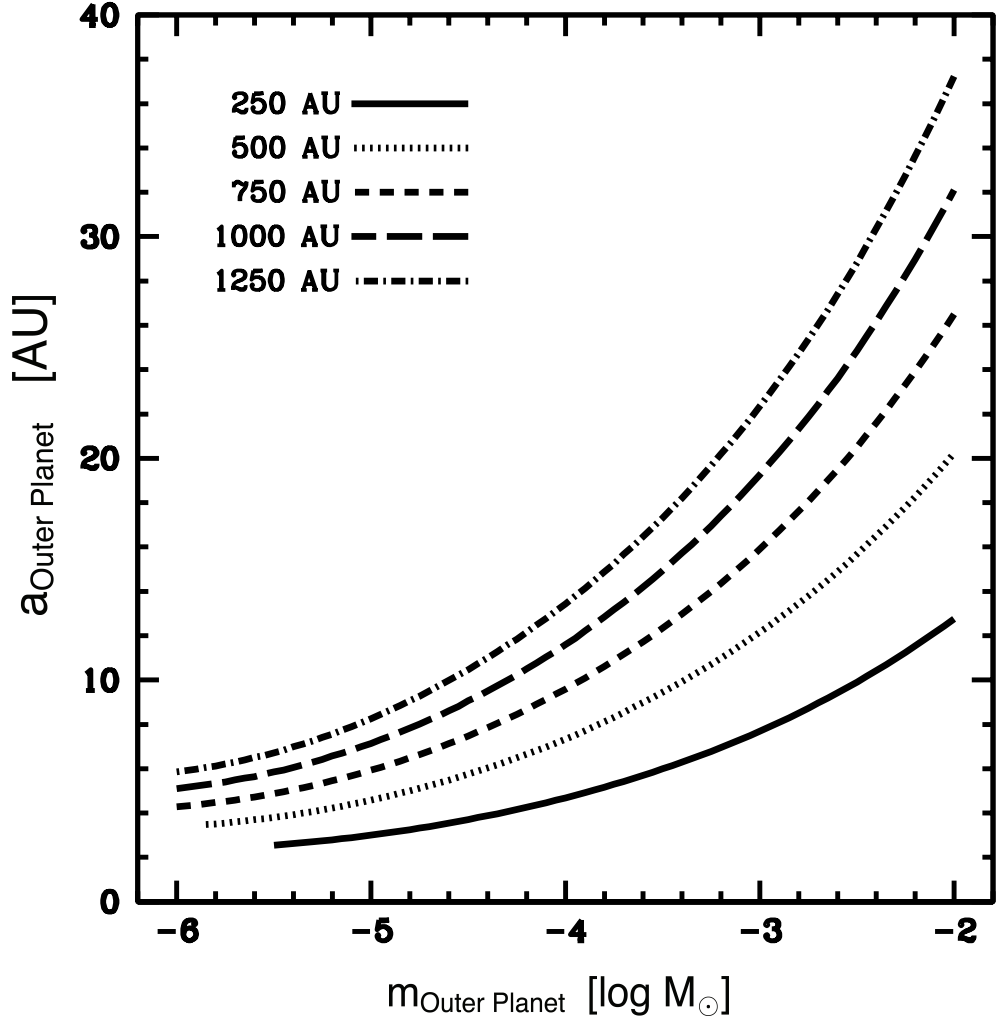}
\end{center}
\caption{ Calculated semimajor axis and mass of the outer planet for which Region D will occur for a system with an earth-like planet.  The data is based on calculations derived from the timescale analysis for various binary-star separations indicated by the line type. 
}
\end{figure}
\clearpage

\begin{figure}
\begin{center}
\includegraphics[width=150mm]{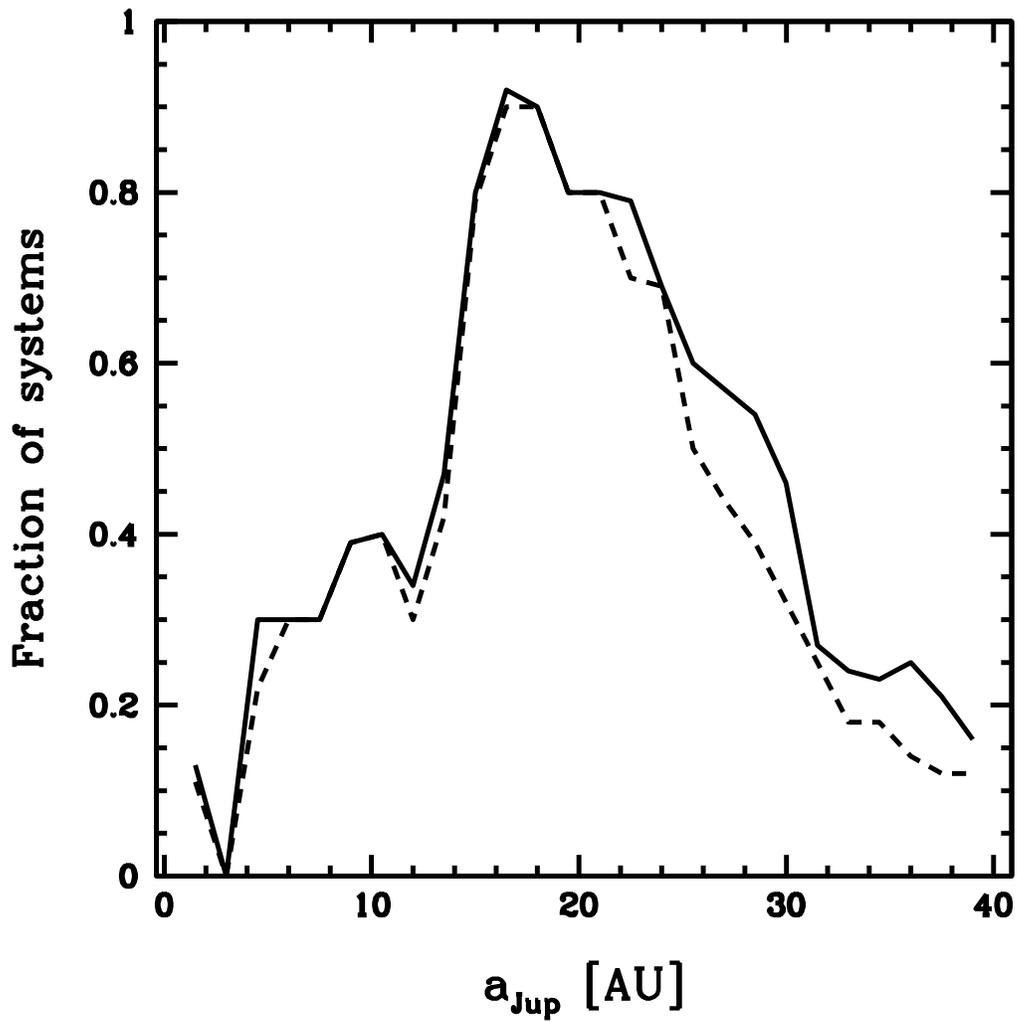}
\end{center}
\caption{ Fraction of systems where the Earth exhibited an eccentricity above 0.4 (dotted) and 0.7 (solid) with respect to $a_{\rm Jup}$, from the simulations performed for Figure 1.  }
\end{figure}

\clearpage

\end{document}